\documentclass[twocolumn]{jpsj3}
\usepackage{graphicx}
\usepackage{txfonts}
\usepackage{bm}
\usepackage{color}

\def\egf{EuGa$_4$}

\def\tn{$T_{\rm N}$}
 \def\ettxt{Eu$T_2$X$_2$}
 \def\mub{${\mu}_{\rm B}$}
 \def\siga{${\sigma}_{\rm a}$}

\title{Magnetic structure of divalent europium compound EuGa$_4$ studied by \\single crystal time-of-flight neutron diffraction}

\author{Takuro Kawasaki$^1$\thanks{takuro.kawasaki@j-parc.jp}, Koji Kaneko$^{1,2}$, Ai Nakamura$^3$\thanks{Present address: Institute for Materials Research, Tohoku University, Oarai, Ibaraki 311-1313, Japan}, Naofumi Aso$^4$, Masato Hedo$^4$, Takao Nakama$^4$, Takashi Ohhara$^1$, Ryoji Kiyanagi$^1$, Kenichi Oikawa$^1$, Itaru Tamura$^1$, Akiko Nakao$^5$, Koji Munakata$^5$, Takayasu Hanashima$^5$ and Yoshichika \=Onuki$^4$}
\inst{$^1$J-PARC Center, Japan Atomic Energy Agency, Tokai, Naka, Ibaraki 319-1195, Japan\\
$^2$Materials Sciences Research Center, Japan Atomic Energy Agency, Tokai, Naka, Ibaraki 319-1195, Japan\\
$^3$Graduate School of Engineering and Science, University of the Ryukyus, Nishihara, Okinawa 903-0213, Japan\\
$^4$Faculty of Science, University of the Ryukyus, Nishihara, Okinawa 903-0213, Japan \\
$^5$Comprehensive Research Organization for Science and Society, Tokai, Ibaraki 319-1106, Japan} 

\abst{
The magnetic structure of the intermetallic compound {\egf} was investigated using single-crystal neutron diffraction with the time-of-flight (TOF) Laue technique on the new diffractometer SENJU at MLF of J-PARC.
In spite of high neutron absorption of Eu, a vast number of diffraction spots were observed without isotope enrichment.
The magnetic reflections were appeared at the positions with the diffraction indices of $h+k+l{\neq}2n$ below 16 K, indicating that the ordering vector is ${\bm q}$=(0~0~0).
Continuous evolution of magnetic reflection intensity below $T_{\rm N}$ follows a squared Brillouin function for $S$=7/2.
By adopting a wavelength-dependent absorption collection, the magnetic structure of {\egf} was revealed that a nearly full magnetic moment of 6.4~{\mub} of Eu lies within the basal plane of the lattice.
The present study reveals a well-localized divalent Eu magnetism in {\egf} and demonstrates a high ability of SENJU to investigate materials with high neutron absorption.
}

\begin{document}
\maketitle

\section{Introduction}
Multiple degrees of freedom of electrons, spin, orbital and charge, and their interplay have a fundamental role on physical properties of materials.
In rare-earth compounds, $f$-electrons generally have trivalent states and are well localized as valence electrons.
Valence instability is seen in the both edge and middle of lanthanide series, namely, Ce, Sm, Eu and Yb intermetallic compounds.
This instability can be tuned by external parameters, such as temperature, pressure and magnetic field. 
In Eu compounds, a valence transition was discovered in EuPd$_2$Si$_2$ with the tetragonal ThCr$_2$Si$_2$-type structure in which the divalent state changed into the trivalent state at the first order transition at 150~K upon cooling.\cite{Sampathkumaran1981}
Valence instability and heavy-electron state are reported in other member of isostructural {\ettxt} as well.\cite{EMLevin1990,Mitsuda2000,Sakurai2003,Hossain2004,Sun2010,Seiro2011,Mitsuda2012,Guritanu2012}
Magnetism of Eu ions has strong and unique dependence on its valence state; a nonmagnetic state with a finite orbital momentum becomes a ground state in a trivalent case, whereas divalent Eu ions carry large spin moment without orbital contribution, $J$=$S$=7/2.
Therefore Eu$T_2X_2$ can be expected to exhibit emergent phenomena which stems from mutual interplay among valence, spin and orbital degrees of freedom.

\begin{figure}[!b]
	\begin{center}
	
		\includegraphics[width=6cm]{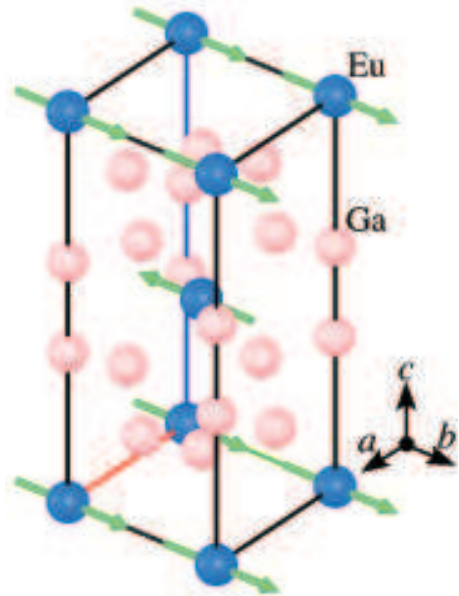}
		\caption{Crystal structure of {\egf}. The antiferromagnetic structure below {\tn} obtained by the present work is also described.}
		\label{struct}
	\end{center}
\end{figure}%

Intermetallic compound {\egf} which has the BaAl$_4$-type crystal structure with the $I4/mmm$ symmetry \cite{Bobev2004} can be regarded as one of the parent variant of {\ettxt} in which both $T$ and $X$ at the 4$d$ and the 4$e$ sites respectively are occupied by Ga.
The crystal structure of  {\egf} is shown in Fig.1.
Recently, high quality single crystals of {\egf} and related compounds have been successfully grown.
Extensive studies revealed that Eu ions in {\egf} have a stable divalent state and hence carry magnetism.
These crystals and the divalent state is ideal to study fundamental magnetic properties of materials under this particular crystal structure. 
{\egf} undergoes a second-order antiferromagnetic transition at {\tn}=16~K.\cite{Nakamura2013}
Antiferromagnetic nature of the transition is identified as a kink in the magnetic susceptibility of the in-plane component, while the $c$-axis component becomes constant below {\tn}.
This behavior is characteristic to a mean-field type antiferromagnet with the ordered moment within the basal plane of the tetragonal lattice.
Validity of the mean-field model with $S$=7/2 is found in other magnetic properties such as isotropic behavior of the paramagnetic susceptibility, and high consistency  between the theoretical magnetic phase diagram and the experimental one.

\begin{figure*}[!t]
	\begin{center}
		\includegraphics[width=13cm]{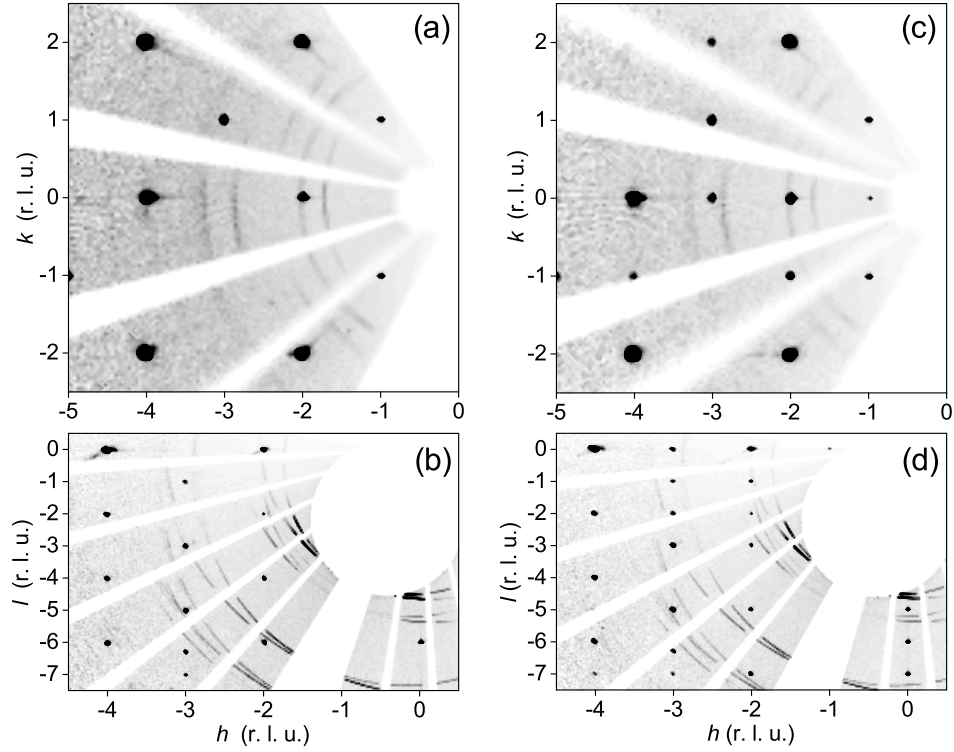}
		\caption{ Neutron diffraction intensities of {\egf} on the ($hk0$) and the ($h0l$) reciprocal lattice planes at 30~K ((a) and (b)),  and 4~K ((c) and (d)). The ring-like intensities correspond to the powder lines from aluminum in the radiation shields of the cryostat.}
		\label{recmap}
	\end{center}
\end{figure*}

In order to understand microscopic magnetic properties of the material, the knowledge on the magnetic structure is indispensable.
In general, neutron diffraction is known as the powerful probe to investigate magnetic structures. 
However, limited numbers of neutron scattering studies have been made on Eu compounds because Eu is known as a strong neutron absorber.
One approach to reduce the absorption effect is isotope enrichment, but this is not easy for Eu due to expensiveness as well as limited availability, in particular in a metallic form.
Another way to confront this difficulty is to choose the wavelength range of neutron that have relatively small absorption cross section for Eu.
Neutron absorption cross sections generally obeys a 1/$v$ law where $v$ is a neutron velocity, in inverse proportional to wavelength ${\lambda}$.
For example, the absorption cross section of 4500~barns at 1.8~{\AA} for natural Eu is suppressed to 880 barns at 0.8~{\AA}, which significantly helps to improve feasibility of neutron diffraction experiments.
Besides, natural Eu shows large absorption for the neutron with the wavelength of around 0.4~{\AA} due to resonance absorption.
In the  time-of-flight (TOF) neutron diffraction experiment at a pulsed neutron source, wide wavelength range neutrons are available and the choice of the specific neutron wavelength range which is suitable for the measurement is possible.

Recently, a single crystal TOF neutron diffractometer SENJU has been built at Materials and Life Science Experimental Facility (MLF) of Japan Proton Accelerator Research Complex (J-PARC).\cite{Ohhara2016}
On SENJU, availability of a wide neutron wavelength range from  0.4~{\AA} to 8.4~{\AA} as well as a large solid angle coverage of 2D position sensitive detectors helps to perform efficient measurement.

In the present study, single crystal neutron diffraction measurements on {\egf} was carried out in order to reveal the magnetic structure in the ground state.

\section{Experimental and analyses}
A single crystal of {\egf} was grown by a Ga self-flux method~\cite{Nakamura2013,NAKAMURA2014}. 
The isotopic abundance of Eu in the present sample is the natural one.

Neutron diffraction experiments were carried out on the single crystal TOF neutron diffractometer SENJU. \cite{Ohhara2016}
The wavelength range of an incident neutron was selected to be 0.4~{\AA}${\sim}$4.4~{\AA} using the bandwidth choppers in the beam line.
The sample with the dimension of ${\sim}3.0{\times}1.6{\times}0.3$~mm$^3$ was glued to a vanadium rod and mounted to a fixed-${\chi}$ two-axes goniometer consists of two piezo rotators.
The goniometer was directly attached to a cold finger of a closed-cycle He refrigerator, which can be cooled down to 4.5~K.
To cover wider area of reciprocal space, the diffraction intensities were measured at several sets of crystal orientation defined by the two rotators.
For structural analysis, intensity data were collected at 2 orientations for 3 hours at 30~K (above {\tn}), and 4 orientations for 4.5 hours at 4.5~K (below {\tn}).
The order parameter was measured for an hour at each temperature and at a fixed orientation from 20~K down to 4.5~K. 
The accelerator power of J-PARC was around 290~kW.
Data reductions and the visualization of the diffraction intensity distributions were performed using the software STARGazer~\cite{Ohhara2009}.

For quantitative analyses, an absorption correction was adopted using the program DABEXN.
Since the software is made for monochromatic data and does not support the correction for pulsed neutron diffraction data, a supplementary Python script was used.
The wavelength-dependent cross section required for the correction was obtained from the Experimental Nuclear Reaction Data (EXFOR)\cite{Otuka2014}. 
A shape of the sample crystal was approximated by a rectangular parallelepiped.
Least square structure refinements were performed using the software JANA2006\cite{Petricek2014}.

\section{Results and Discussions}
A sufficient number of Bragg reflections were observed in the measurement.
The neutron diffraction intensity distributions on the ($h0l$) and the ($hk0$) reciprocal lattice planes at 30~K and 4.5~K are shown in Fig.\ref{recmap}.
The lattice parameters were obtained at $a=b=$4.3889(4)~{\AA} and $c=$10.6479(5)~{\AA} for 30 K, and $a=b=$4.3795(3)~{\AA} and $c=$10.6302(7)~{\AA} for 4.5 K, respectively, 
which are comparable to the reported values\cite{Bobev2004,Nakamura2013}. 
At 30~K above {\tn}, nuclear Bragg spots were observed under the condition of $h+k+l$=2$n$ ($n$: integer), which follows the extinction rule of $I4/mmm$ symmetry.
In the antiferromagnetic state at 4.5~K below {\tn}, superlattice reflections were observed at the positions of $h+k+l{\neq}2n$.
The intensities of the superlattice reflections decreased as the momentum transfer $Q$ increases, being indicative of their magnetic origin.
The integer diffraction indices $hkl$ of the magnetic reflections reveals that the antiferromagnetic structure of {\egf} is described with the propagation vector $\bm{q}$=(0~0~0) below {\tn}. 
Further, the violation of the extinction condition means body-centered translation was broken in the magnetic structure of {\egf}.

\begin{figure}[!t]
	\begin{center}
		\includegraphics[width=8cm]{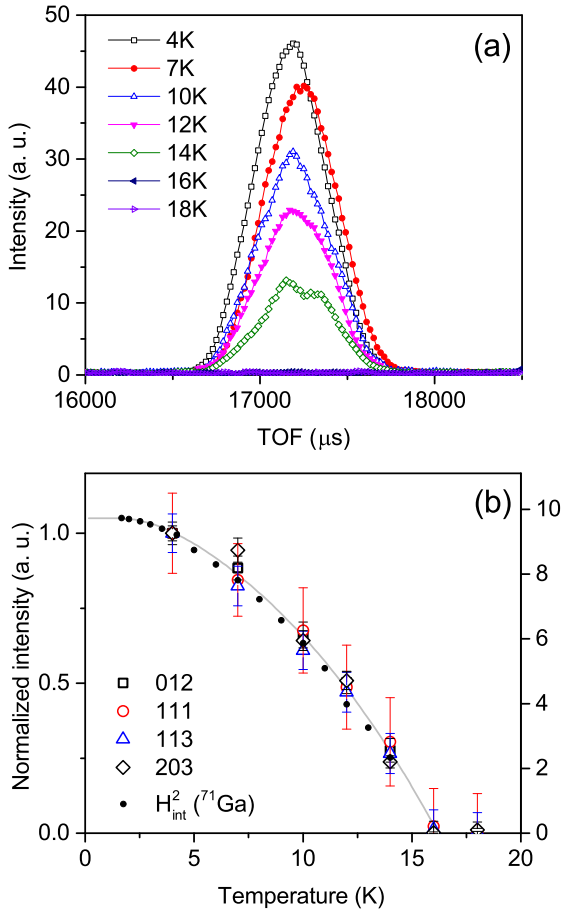}
		\caption{ (a) The peak profile of the 012 reflection around and below {\tn}, and (b) the temperature dependence of the integrated intensity of magnetic superlattice reflections. 
		A square of the internal field at Ga site\cite{Yogi2013} are plotted as a function of temperature in the same figure. The line in this figure is a squared Brillouin function for $S$=7/2.}
		\label{temperature}
	\end{center}
\end{figure}

Figure~\ref{temperature}(a) shows the temperature dependences of the diffraction profile of the 012 reflection. 
The superlattice reflection continuously develops as the temperature decreases below {\tn}.
The temperature evolution of the integrated intensities for several superlattice reflections is plotted in Fig.~\ref{temperature}(b), which are normalized by the intensity at 4.5~K.
The temperature variations of each integrated intensity are identical to each other. Since the $Q$-vectors of each reflection are different, a continuous evolution of the magnetic ordering without reorientation of the magnetic moment is suggested below {\tn}.
In the same figure, the internal magnetic field measured at the $^{71}$Ga site and the scaled Brillouin function for $S$=7/2 taken from ref. 17 are displayed in the right axis in squared form, compare with the neutron intensity proportional to ${\mu}^2$.\cite{Yogi2013}
A remarkable agreement of the temperature dependence is seen in quantities obtained from the independent measurements.
This clearly demonstrates that the magnetic transition at {\tn} in {\egf} has a mean-field type second-order nature.

\begin{figure}[!t]
	\begin{center}
		\includegraphics[width=8.5cm]{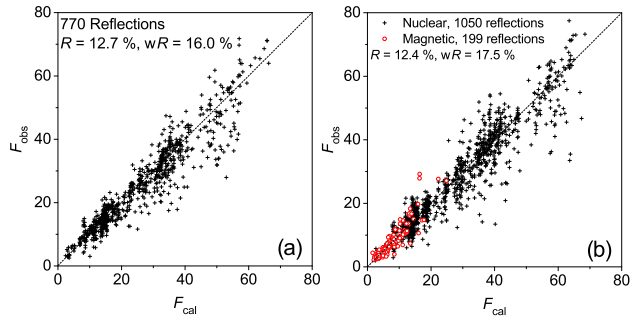}
		\caption{The observed and the calculated structure factors of {\egf} (a) at 30~K and (b) at 4.5~K.}
		\label{Fcalobs}
	\end{center}
\end{figure}

Hereafter, we describe a quantitative part of the analysis.
At first, the crystal structure refinement was performed for the data recorded at 30~K. 
In the following analyses, the intensity data corresponding to the wavelength range between 0.5${\sim}$2.0~{\AA} are used because of the relatively smooth, and reduced absorption cross section of Eu where {\siga}~$<{\sim}$5000~barns, and to avoid resonance absorption.
Based on the reported crystal structure, 7 parameters that consists of the fractional coordinate $z$ for Ga2 at the 4$e$ site, 
isotropic displacement parameters for each element, the scale factors for two crystal orientation data sets and the extinction parameter, were refined 
using 312 independent reflections out of 770 reflections which satisfy $I>5{\sigma}I$ where ${\sigma}I$ corresponds to an experimental error for each reflection.
The observed structure factors ($F_{\rm obs}$) are plotted against the calculated ones ($F_{\rm cal}$) for 30 K in Fig.~\ref{Fcalobs}.
The refined structural parameters listed in Table I are agree with the previous ones,  with the reasonable reliable factors of $R$=12.7~\% and $wR$=16.0~\%. 
This result confirms the validity of the present absorption correction and therefore the same corrections are applied for the subsequent analyses for low temperature data.

The simultaneous crystal and magnetic structure analyses were performed for the intensities measured at 4.5~K. 
Concerning the magnetic structure, the magnetic structure model is assumed as in Fig.~\ref{struct} for the following reasons;
(i)~The observed propagation vector of ${\bm q}=(0~0~0)$ indicates that magnetic moments at the corner and the body-center positions are antiparallel.
(ii)~Magnetization measurements implies that the moment lies within the basal $ab$-plane, maybe along the ${\langle}1~0~0{\rangle}$ direction and forms domains.
The magnetic form factor of Eu was assumed to be Eu$^{2+}{\langle}j_0{\rangle}$ in the refinement. 

\begin{figure}[!t]
	\begin{center}
		\includegraphics[width=8cm]{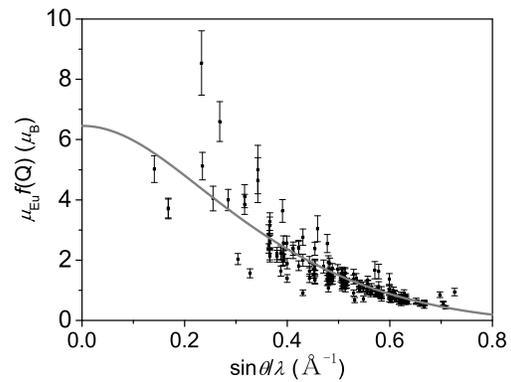}
		\caption{$sin{\theta}/{\lambda}$ dependence of the magnetic form factor of Eu in {\egf} at 4.5~K. The points represent measured values, and the line represent calculated curve based on the equation and the parameters in ref.\cite{ITC}, and the magnetic moment of Eu = 6.4~{\mub} obtained from the present analysis.}
		\label{formfact}
	\end{center}
\end{figure}

Including two additional parameters, the amplitude of the moment and the twin fraction, for this magnetic structure, 9 parameters in total were refined using 1050 nuclear and 199 magnetic reflections. 
The satisfactory reliable factors of $R$=12.4~\% and $wR$=~17.5~\% comparable to those for 30 K were reached.
The refined parameters are listed in table II. 
The deduced magnetic moment of Eu at 4.5 K of 6.4(2) {\mub}, is close to that of Eu$^{2+}$ of 7 {\mub}. 
The refined twin fractions of  0.54(6) : 0.46(6) indicating the equal domain population within the error is reasonable. 
Validity of the magnetic form factor $f$({\boldmath$Q$}) of Eu$^{2+}$ is examined by the transformation of the intensity data $I_{\rm mag}$ using a following equation,
\begin{equation}
	{\mu} f(\mbox{\boldmath$Q$}) \propto ( I_{\rm mag}({hkl}) / L (\theta))^{1/2}/( S_{\perp}^{hkl}{\mid}F_{\rm mag}({hkl}){\mid}) 
	\label{eq:Nint}
\end{equation}
where $L({\theta})$ and $F_{\rm mag}$ are the Lorentz and the magnetic structure factors, and $S_{\perp}^{hkl}$ is the magnetic interaction vector, which reflects that neutron scattering observes only magnetic moment components perpendicular to the momentum transfer {\boldmath$Q$}.
Figure~\ref{formfact} shows the derived product of ${\mu} f(\bm{Q}$) of Eu as a function of ${\sin}~{\theta}/{\lambda}$. 
The experimentally observed values well coincide with the calculated ones based only on $j$$_{0}$ for Eu$^{2+}$ with ${\mu}$=6.4(2)~{\mub}, which also confirms a divalent $S$=7/2 magnetism in {\egf}. 
While a better agreement between the experimental data and the calculation is obtained in ${\sin}~{\theta}/{\lambda}>0.35$, a discrepancy is prominent in lower momentum transfer.
Because of these data were measured with longer wavelength neutron and therefore strongly affected by the absorption correction.

\begin{table}
\caption{Refined structural parameters of {\egf} at 30 K.}
\label{t1}
\begin{center}
\begin{tabular}{llllll}
\hline
\verb|Atom| & Wyckoff & x & y & z & $U_{iso}$ \\
\hline
\verb|Eu| & 2a & 0 & 0 & 0 & 0.0022(2) \\
\verb|Ga1| & 4d & 0 & 1/2 & 1/4 & 0.0006(2) \\
\verb|Ga2| & 4e & 0 & 0 & 0.3838(1) & 0.0002(2) \\
\hline
\end{tabular}
\end{center}
\end{table}

\begin{table}
\caption{Refined structural parameters of {\egf} at 4.5 K.}
\label{t2}
\begin{center}
\begin{tabular}{lllllll}
\hline
\verb|Atom| & Wyckoff & x & y & z & $U_{iso}$ & M({\mub})\\
\hline
\verb|Eu| & 2a & 0 & 0 & 0 & 0.0016(3) & 6.4(2)\\
\verb|Ga1| & 4d & 0 & 1/2 & 1/4 & 0.0002(2) \\
\verb|Ga2| & 4e & 0 & 0 & 0.3836(1) & 0.0003(2) \\
\hline
\end{tabular}
\end{center}
\end{table}

The present neutron diffraction experiment successfully revealed the antiferromagnetic structure of {\egf} below {\tn}.
A remarkable agreement in the evolution of the order parameter among independent measurements, neutron diffraction, NMR together with M\"ossbauer spectroscopy\cite{Homma2014} which follows the Brillouin function for $S$=7/2 was observed. 
Furthermore, the obtained magnetic form factor of Eu$^{2+}$ and the isotropic magnetic phase diagram confirm the spin only magnetism with  $S$=7/2.
Namely, these experimental data demonstrates that {\egf} is an ideal mean-field type antiferromagnet with the stable divalent Eu ion.

Whereas the magnetic structure was revealed in the sample without isotope enrichment, the quantitative agreement is moderate.
A large discrepancy is seen, in particular, at the low momentum transfer in Fig.~\ref{formfact} where an effect of absorption correction is enormous due to longer wavelength.
In order to achieve better reliable factor, accurate information on a sample shape should be crucial, since the attenuation length, with which the incident neutron flux is reduced to 1/$e$, is of the order of a few tenth of micrometer for Eu.
A dedicated equipment is necessary for this aim.
The accuracy in the energy dependent absorption cross section is required as well.

On the other hand, this work demonstrates that the TOF neutron diffraction experiment on SENJU can serve as a powerful tool to determine magnetic propagation vectors of Eu compounds without isotope enrichment.
Due to the strong absorption, a small single crystal with the 1~mm$^3$ in size is sufficient for this type of experiment. 
A complementary use of microscopic probes, such as neutron scattering, NMR and M\"ossbauer enables us to gain quantitative insights into magnetic ordering.  
We plan to extend our neutron scattering study on Eu compounds for intermediate valence systems as well as to apply inelastic spectroscopy by taking advantage of recent progress on both neutron source and instrumentation.

\section{Summary}

The magnetic structure of {\egf} was investigated by means of TOF neutron diffraction without isotope enrichment for Eu.
The magnetic structure is characterized with ${\bm q}$=(0~0~0) where the magnetic moments lie within the basal $ab$-plane.
The crystal and magnetic structural analyses with the energy dependent absorption correction works satisfactory, which gives a reasonable size of Eu moment of 6.4~{\mub} at 4.5~K. 
The present study reveals a well-localized divalent Eu magnetism in {\egf} and demonstrates effectiveness of SENJU to study materials with strong neutron absorption.

We would like to thank M. Yogi, S. Shimomura and Y. Homma for valuable discussions. 
This work was partly supported by Grant-in-Aids for Scientific Research (C) (No. 24540336, No. 26400348) from Japan Society for the Promotion of Science.
The measurement was performed using SENJU along with the project use program (2012P0203) and the general use program (2014A0080) of J-PARC and CROSS.
%

\bibliographystyle{jpsj}

\begin{thebibliography}{10}

\bibitem{Sampathkumaran1981}
E.~V. Sampathkumaran, L.~C. Gupta, R.~Vijayaraghavan, K.~V. Gopalakrishnan,
  R.~G. Pillay, and H.~G. Devare: J. Phys. C Solid State Phys. {\bfseries 14}
  (1981) L237 .

\bibitem{EMLevin1990}
E.~M. Levin, B.~S. Kuzhel, O.~I. Bodak, B.~D. Belan, and I.~N. Stets: Phys.
  Stat. Sol. {\bfseries 783} (1990).

\bibitem{Mitsuda2000}
A.~Mitsuda, H.~Wada, M.~Shiga, and T.~Tanaka: J. Phys. Condens. Matter
  {\bfseries 12} (2000) 5287.

\bibitem{Sakurai2003}
J.~Sakurai, Y.~Nakanuma, S.~Fukuda, A.~Mitsuda, and Y.~Isikawa: J. Phys. Soc.
  Jpn. {\bfseries 72} (2003) 2046.

\bibitem{Hossain2004}
Z.~Hossain, C.~Geibel, N.~Senthilkumaran, M.~Deppe, M.~Baenitz, F.~Schiller,
  and S.~Molodtsov: Phys. Rev. B {\bfseries 69} (2004) 014422 1.

\bibitem{Sun2010}
L.~Sun, J.~Guo, G.~Chen, X.~Chen, X.~Dong, W.~Lu, C.~Zhang, Z.~Jiang, Y.~Zou,
  S.~Zhang, Y.~Huang, Q.~Wu, X.~Dai, Y.~Li, J.~Liu, and Z.~Zhao: Phys. Rev. B
  {\bfseries 82} (2010) 4.

\bibitem{Seiro2011}
S.~Seiro and C.~Geibel: J. Phys. Condens. Matter {\bfseries 23} (2011) 375601.

\bibitem{Mitsuda2012}
A.~Mitsuda, S.~Hamano, N.~Araoka, H.~Yayama, and H.~Wada: J. Phys. Soc. Jpn.
  {\bfseries 81} (2012) 023709.

\bibitem{Guritanu2012}
V.~Guritanu, S.~Seiro, J.~Sichelschmidt, N.~Caroca-Canales, T.~Iizuka,
  S.~Kimura, C.~Geibel, and F.~Steglich: Phys. Rev. Lett. {\bfseries 109}
  (2012) 247207.

\bibitem{Bobev2004}
S.~Bobev, E.~D. Bauer, J.~D. Thompson, and J.~L. Sarrao: J. Magn. Magn. Mater.
  {\bfseries 277} (2004) 236.

\bibitem{Nakamura2013}
A.~Nakamura, Y.~Hiranaka, M.~Hedo, T.~Nakama, Y.~Miura, H.~Tsutsumi, A.~Mori,
  K.~Ishida, K.~Mitamura, Y.~Hirose, K.~Sugiyama, F.~Honda, R.~Settai,
  T.~Takeuchi, M.~Hagiwara, T.~D. Matsuda, E.~Yamamoto, Y.~Haga,
  K.~Matsubayashi, Y.~Uwatoko, H.~Harima, and Y.~\=Onuki: J. Phys. Soc. Jpn.
  {\bfseries 82} (2013) 104703 1.

\bibitem{Ohhara2016}
T.~Ohhara, R.~Kiyanagi, K.~Oikawa, K.~Kaneko, T.~Kawasaki, I.~Tamura, A.~Nakao,
  T.~Hanashima, K.~Munakata, T.~Moyoshi, T.~Kuroda, H.~Kimura, T.~Sakakura,
  C.-H. Lee, M.~Takahashi, K.-i. Ohshima, T.~Kiyotani, Y.~Noda, and M.~Arai: J.
  Appl. Crystallogr. {\bfseries 49} (2016) 1.

\bibitem{NAKAMURA2014}
A.~Nakamura, Y.~Hiranaka, M.~Hedo, T.~Nakama, Y.~Miura, H.~Tsutsumi, A.~Mori,
  K.~Ishida, K.~Mitamura, Y.~Hirose, K.~Sugiyama, F.~Honda, T.~Takeuchi, T.~D.
  Matsuda, E.~Yamamoto, Y.~Haga, and Y.~\=Onuki: JPS Conf. Proc. {\bfseries 3}
  (2014) 011012.

\bibitem{Ohhara2009}
T.~Ohhara, K.~Kusaka, T.~Hosoya, K.~Kurihara, K.~Tomoyori, N.~Niimura,
  I.~Tanaka, J.~Suzuki, T.~Nakatani, T.~Otomo, S.~Matsuoka, K.~Tomita,
  Y.~Nishimaki, T.~Ajima, and S.~Ryufuku: Nucl. Instruments Methods Phys. Res.
  Sect. A {\bfseries 600} (2009) 195.

\bibitem{Otuka2014}
N.~Otuka, E.~Dupont, V.~Semkova, B.~Pritychenko, A.~I. Blokhin, M.~Aikawa,
  S.~Babykina, M.~Bossant, G.~Chen, S.~Dunaeva, R.~A. Forrest, T.~Fukahori,
  N.~Furutachi, S.~Ganesan, Z.~Ge, O.~O. Gritzay, M.~Herman, S.~Hlava{\v{c}},
  K.~Kato, B.~Lalremruata, Y.~O. Lee, A.~Makinaga, K.~Matsumoto,
  M.~Mikhaylyukova, G.~Pikulina, V.~G. Pronyaev, A.~Saxena, O.~Schwerer, S.~P.
  Simakov, N.~Soppera, R.~Suzuki, S.~Tak{\'{a}}cs, X.~Tao, S.~Taova,
  F.~T{\'{a}}rk{\'{a}}nyi, V.~V. Varlamov, J.~Wang, S.~C. Yang, V.~Zerkin, and
  Y.~Zhuang: Nucl. Data Sheets {\bfseries 120} (2014) 272.

\bibitem{Petricek2014}
V.~Pet\v{r}\'{i}\v{c}ek, M.~Du\v{s}ek, and L.~Palatinus: Z. Kristallogr.
  {\bfseries 229} (2014) 345.

\bibitem{Yogi2013}
M.~Yogi, S.~Nakamura, N.~Higa, H.~Niki, Y.~Hirose, Y.~\=Onuki, and H.~Harima:
  J. Phys. Soc. Jpn. {\bfseries 82} (2013) 103701 1.

\bibitem{ITC}
{\em {International Tables for Crystallography Volume C: Mathematical, physical
  and chemical tables}}, ed. E.~Prince (Kluwer Academic Publishers, 2004).

\bibitem{Homma2014}
Y.~Homma, A.~Nakamura, Y.~Hirose, M.~Hedo, and T.~Nakama: JPS Conf. Proc.
  {\bfseries 4} (2014) 16.

\end{thebibliography}

\end{document}